\documentclass[a5paper,graybox]{svmult} 

\usepackage[round]{natbib}
\usepackage{type1cm}        
\usepackage{makeidx}        
\usepackage{graphicx}       
\usepackage{multicol}       
\usepackage[bottom]{footmisc}
\usepackage{xcolor}         
\usepackage{hyperref}

\definecolor{mylinkcolor}{HTML}{9a9171}
\definecolor{mylinkcolor2}{HTML}{f8f8f7}
\hypersetup{
    colorlinks=true,
    linkcolor=mylinkcolor,  
    citecolor=mylinkcolor,  
    urlcolor=mylinkcolor,   
}
\makeatletter
\renewcommand\@cite[2]{%
    [{\hypersetup{citecolor=mylinkcolor}[{#1\if@tempswa , #2\fi}]}]
}
\makeatother

\usepackage{newtxtext} 
\usepackage{newtxmath}

\usepackage{geometry}
\graphicspath{{Figures/}}

\usepackage{caption}

\usepackage{parskip}

\usepackage{background}  

\backgroundsetup{
  scale=4.75,  
  color=black,  
  opacity=0.1,  
  angle=0,  
  vshift=70pt,  
  contents={\includegraphics{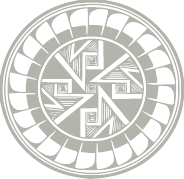}}  
}

\usepackage{tikz,xcolor,hyperref}
\definecolor{lime}{HTML}{A6CE39}
\DeclareRobustCommand{\orcidicon}{%
    \begin{tikzpicture}
    \draw[mylinkcolor, fill=mylinkcolor] (0,0) 
    circle [radius=0.16] 
    node[white] {{\fontfamily{qag}\selectfont \tiny ID}};
    \draw[white, fill=white] (-0.0625,0.095) 
    circle [radius=0.007];
    \end{tikzpicture}
    \hspace{-2mm}
}
\newcommand{\orcid}[1]{\href{https://orcid.org/#1}{\orcidicon}}

\usepackage{everypage}  
\AddEverypageHook{%
  \begin{tikzpicture}[remember picture, overlay]
    \node[anchor=south, xshift=0mm, yshift=1mm] at (current page.south) {
      \begin{tikzpicture}
        \node[fill=mylinkcolor2, text=black, rounded corners=2mm, inner sep=3mm]{
          \footnotesize \textit{This article will be a contributed chapter to the SFI edited volume: The Economy as a Complex Evolving System, Part IV.}
        };
      \end{tikzpicture}
    };
  \end{tikzpicture}%
}

\begin{document}

\title*{The Evolutionary Ecology of Software: Constraints, Innovation, and the AI Disruption}
\author{Sergi Valverde, Blai Vidiella, and Salva Duran-Nebreda }

\institute {
Sergi Valverde \at Institute of Evolutionary Biology (CSIC-UPF)\\ \email{s.valverde@csic.es}
\and
Blai Vidiella \at Theoretical and Experimental Ecology Station (CNRS) \\ \email{blaivr.lab@gmail.com}
\and
Salva Duran-Nebreda \at Institute of Evolutionary Biology (CSIC-UPF) \\ \email{salva.duran@ibe.upf-csic.es}
}

\maketitle

\abstract{This chapter investigates the evolutionary ecology of software, focusing on the symbiotic relationship between software and innovation. An interplay between constraints, tinkering, and frequency-dependent selection drives the complex evolutionary trajectories of these socio-technological systems. Our approach integrates agent-based modeling and case studies, drawing on complex network analysis and evolutionary theory to explore how software evolves under the competing forces of novelty generation and imitation. By examining the evolution of programming languages and their impact on developer practices, we illustrate how technological artifacts co-evolve with and shape societal norms, cultural dynamics, and human interactions. This ecological perspective also informs our analysis of the emerging role of AI-driven development tools in software evolution. While large language models (LLMs) provide unprecedented access to information, their widespread adoption introduces new evolutionary pressures that may contribute to cultural stagnation, much like the decline of diversity in past software ecosystems. Understanding the evolutionary pressures introduced by AI-mediated software production is critical for anticipating broader patterns of cultural change, technological adaptation, and the future of software innovation.}


\section*{Introduction}\label{sec:intro}

Since the dawn of the digital age, there has been a strong mutual influence between biologists and computer scientists. Computer scientists have attempted to model computers after biological processes using a variety of approaches, including cellular automata, neural networks, and evolutionary algorithms. In contrast, biologists have used computational metaphors to illustrate how biological systems operate. For example, geneticists have linked DNA to software, and neuroscientists have claimed that brain functions are similar to those performed by computers. 

In spite of this mutual influence, there is a palpable, sharp divide between biology and computer science. Evolutionary algorithms, like genetic algorithms, are inspired by organismic evolution, although they have evident limitations in terms of complexity and open-endedness. On the other hand, biologists opposed computational analogies as early as 1951, when neuroscientist Karl Lashley argued that ``we are more likely to find out how the brain works by studying the brain itself, and the phenomena of behaviour, than by indulging in far-fetched physical analogies~\cite{beach1960neuropsychology}.''

To deepen the debate about the differences and similarities between software and biology, it is crucial to move beyond the superficial assumption that technology is different from living systems mainly because it is built by humans. Several examples, such as the present climate crisis, demonstrate the limits of technological planning. This is relevant when projecting the long-term uses of technology, as advances in software are not always the result of well-planned, dependable, and managed systems. Instead, all technologies evolve via a cumulative process based on previous, empirically-tested practices~\cite{richerson2008not}. Software is a well-documented example of cultural evolution, which is the study of how ideas, behaviors, technologies, and practices change, spread, and adapt within societies over time~\cite{vidiella2022cultural, sole2013evolutionary}.  

We will examine the evolutionary and ecological drivers of software innovation. Our methodology departs from previous approaches that depend on qualitative biological metaphors, since we adopt the quantitative techniques of empirical and theoretical network analysis. 
With this, we do not aim to demonstrate that software and biology are the same, but rather to highlight that they can be analyzed using similar tools. As society becomes more dependent on software, it is critical to develop a technologically oriented evolutionary approach capable of identifying the causes of constraints, enabling collaborative problem-solving and facilitating technological advances. 

\begin{figure}[hb!]
    \centering
    \includegraphics[width=0.85\textwidth]{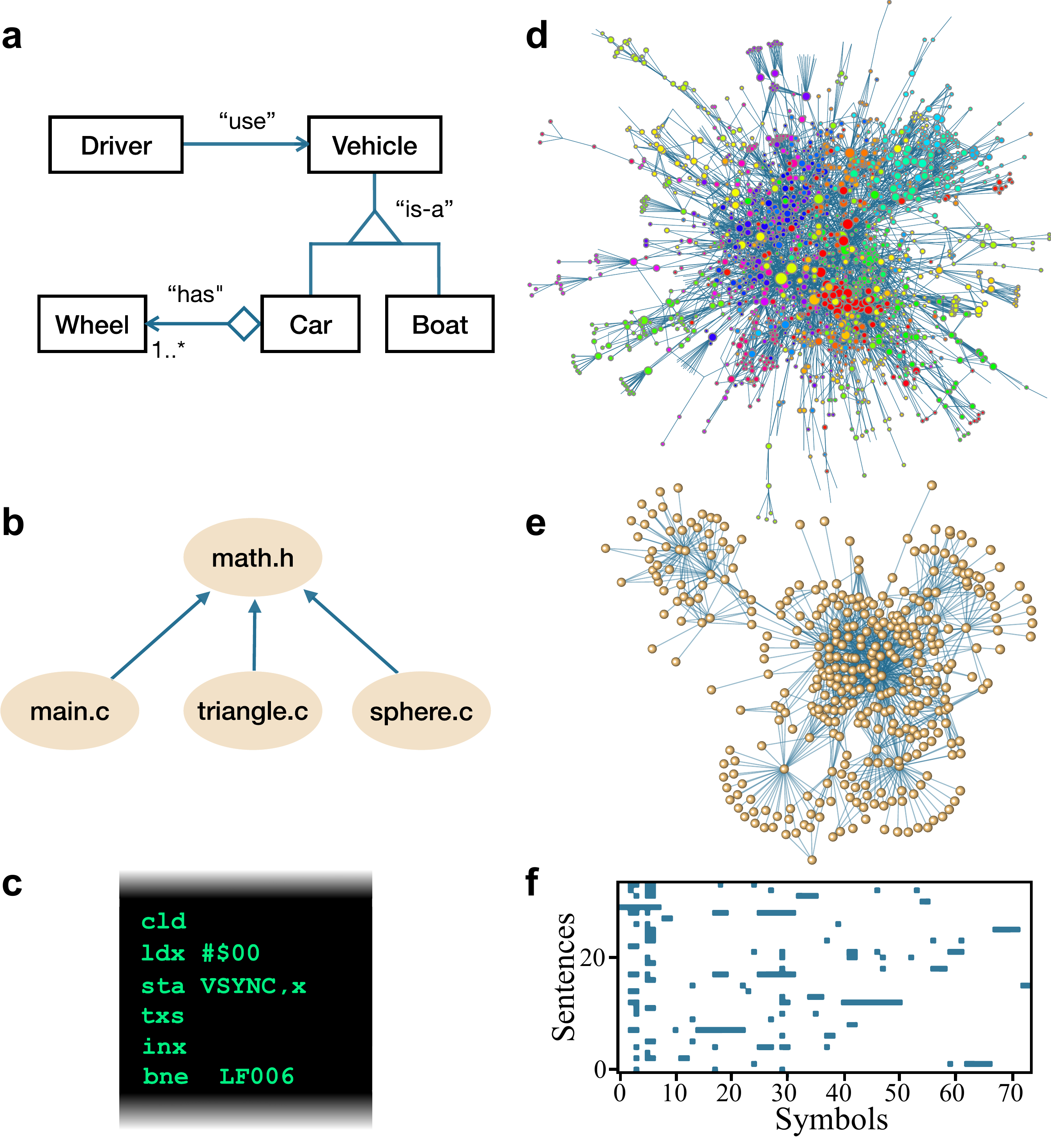}
    \captionof{figure}{{\bf Network representations of software systems.} {\bf (a)} Software engineers use class diagrams, a type of semantic map described in unified modeling language (UML) notation, to organize software components and their interactions. We can see part of the class diagram used to describe the elements in a racing game, including several types of vehicles (``is-a'' relationships) and the association between a car and its wheels (a ``has'' relationship). {\bf (b)} A dependency network distributes software functions into a directed acyclic graph of source code files and their dependencies on other code files. ...}
    \label{fig:1}
\end{figure}

\begin{figure}[ht!]
    \ContinuedFloat
    \caption[]{
    ... {\bf (c)} Source code is a text file containing the software instructions written in a language of machine symbols, including arithmetic operations, register names, and memory addresses. {\bf (d)} Software network for the class diagram used in the development of the console game Pro Rally 2002 (Ubisoft, 2002). Color denotes the subsystem the software component belongs to (i.e., physics, rendering, audio, etc.), and node size represents the in-degree. {\bf (e)} The dependency network of the XFree86 X Window System exhibits scale-free behavior (code files from the version published on May 15, 1993). {\bf (f)} Every network may be represented using an adjacency matrix, including the bipartite network that shows the relationship between sentences and machine symbols in source code.}
\end{figure}

\section*{Software Networks Are Scale-Free}

Software is an integral part of our daily life, yet it often introduces unexpected complexity, even though software engineers are constantly refining system design and structure, providing ``patches'' that fix any flaws ~\cite{valverde2021long}. One of the main goals for software designers is to create systems that are easier to understand, handle, and update. To accomplish this, they divide the system into components with predictable interactions while adhering to modular and hierarchical design principles ~\cite{baldwin2000design}. This approach has worked well in electronics because understanding the circuit as a whole does not necessitate understanding the internal details of the underlying components. However, why aren't these principles fully applicable to software?

The outputs of planned design must be considered highly optimized systems~\cite{valverde2002scale}, although design goals might come into conflict with the practical requirements of dealing with environmental complexities, which are often mirrored in the internal structure of technological networks~\cite{mcnerney2011role}. To measure the level of structural complexity, we examine the software network represented as a graph $G=(V,E)$, where $V$ denotes the software elements (such as classes, files, or machine symbols) and connections $(i,j) \in E$ depict semantic relationships among classes (see fig.~\ref{fig:1}a), dependencies between source code files (see fig.~\ref{fig:1}b), or associations between symbols and sentences within source code files (see fig.~\ref{fig:1}c). 

Many real-world networks, including software networks, have a structure that is highly heterogeneous (fig.~\ref{fig:1}d), with a small number of hub nodes having many connections, while the majority have only a few links.  For example, in software dependency networks, the probability that a random node has $k$ incoming links (or in-degree distribution) follows a power-law (see fig.~\ref{fig:2}a): 
\begin{equation*}
P(k) \sim k^{-\gamma},
\end{equation*}
where the exponent $\gamma \approx 2$~\cite{valverde2005logarithmic, de2009analysis}. Similar scale-free degree distributions are seen in class diagrams as well ~\cite{valverde2002scale, valverde2007hierarchy}. This heterogeneous organization, which is not intentionally imposed by software engineers, significantly impacts the development of software systems~\cite{lakos1996large}, suggesting that we must work around complexity constraints to successfully design and manage these systems. 

These heterogeneous degree distributions, a signature of emergent complexity, may not capture all aspects of software design. Engineers integrate multiple components to develop functionalities that are too complex to be contained within a single piece of software. For example, a ``design pattern'' is a template that includes interacting software parts that may be reused to address common design problems~\cite{gamma1995design}. Quantitative data on the structural aspects of design patterns, such as the frequency of software subgraphs (see fig.~\ref{fig:2}b), may aid in establishing their utility, as it has been proposed for regulatory networks, where specific subgraphs (or network motifs) represent information-processing building blocks~\cite{milo2002motifs}. The importance of subgraph abundances, however, is unknown, since identical motifs may perform various functions in different systems~\cite{knabe2008motifs}, demonstrating how biological and software components cannot be totally isolated from contextual factors. To disentangle these factors, a proper null model is needed.

\section*{Software Evolves by Tinkering}

In software development, ``tinkering,'' or playful, unguided learning, refers to the process of experimenting with code. An informal and iterative approach to learning promotes curiosity and flexibility, often leading to unexpected discoveries and innovative solutions~\cite{martocchio1992effects,webster1993turning}. Tinkering (also known as ``bricolage''~\cite{turkle1992epistemological}) relies on experimenting with new combinations (and re-combinations) of existing software elements. 

A tinkering-based model can predict the abundances of motifs in software networks~\cite{valverde2005motifs}, challenging the conventional assumption that functionality is the main driver of design processes~\cite{sole2006motifs, brunswicker2023microstructure}. This network model, which is not reliant on any specific function, is able to accurately predict the evolution of average degree over time and the asymmetrical patterns in the distribution of degrees. 

\begin{figure}[ht!]
    \centering
    \includegraphics[width=0.9\textwidth]{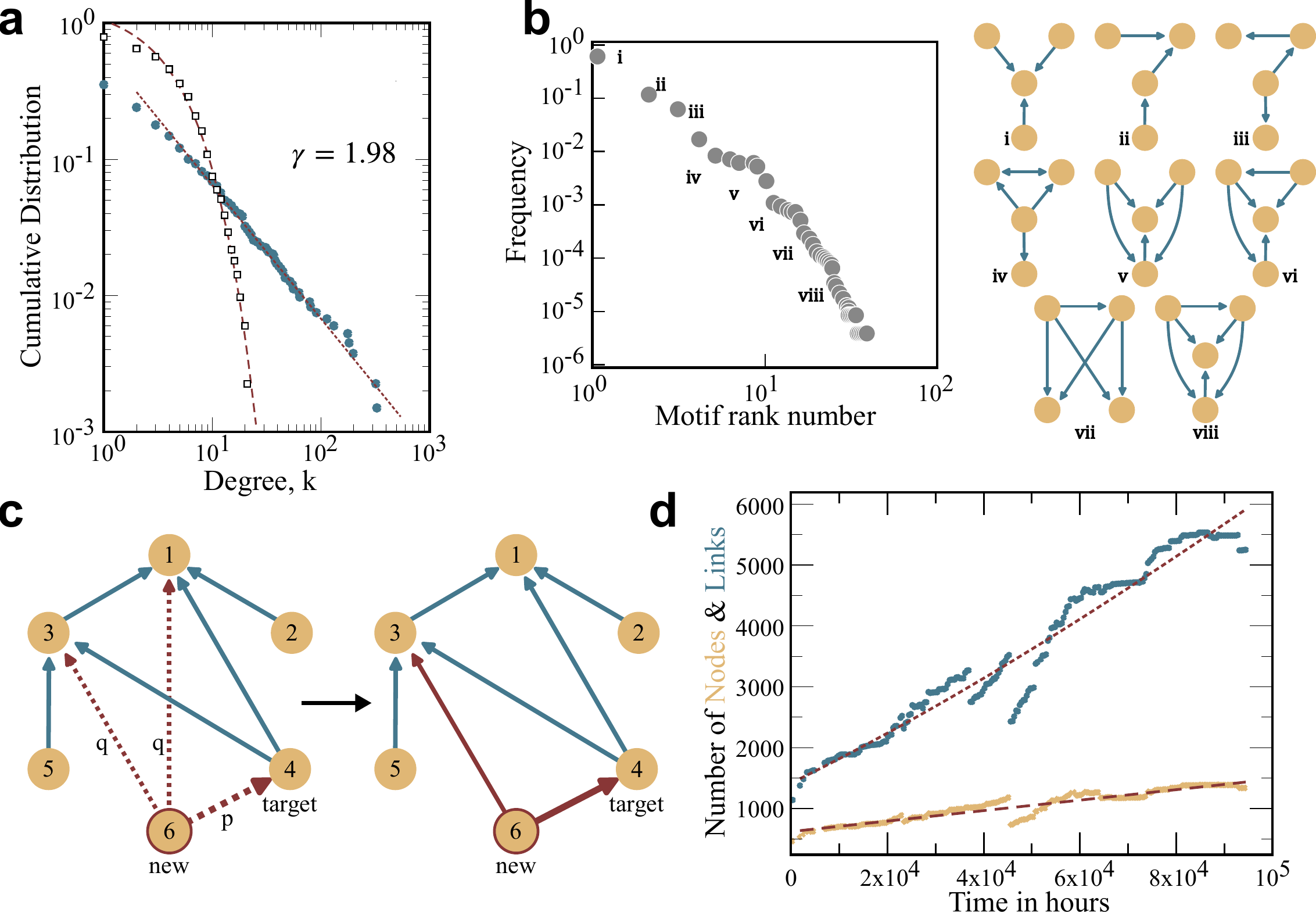}
    \caption{{\bf Tinkered evolution of software networks.} {\bf (a)} Cumulative in-degree (filled circles) and out-degree (white squares) distributions for the project XFree86. The power-law fit of the in-degree distribution yields $P_>(k) \sim k^{-\gamma+1}$ with $\gamma = 1.97 \pm 0.01$ while the out-degree distribution is exponential~\cite{valverde2005logarithmic}. {\bf (b)} Frequency-rank distribution of 4-subgraphs in the software network for the project Exult~\cite{valverde2005motifs}. The frequency $P(R)$ decays rapidly with subgraph rank $k$ indicating how common subgraphs (like i or ii) are sparser than less common ones, which are more dense (like vii or viii). {\bf (c)} Copying rules used in the model of software network growth. Each node is labeled with an age (number 1 is the oldest). A new node $v_6$ links to target node $v_4$ with probability $p$. This new node inherits every connection from the target node (dashed links) with probability $q$. This model predicts topological characteristics of software networks, such as the exponent $\gamma=2$ of power-law in-degree distributions and observed subgraph frequencies~\cite{sole2006motifs}, as well as {\bf (d)} the evolutionary pattern of logarithmic scaling in software~\cite{valverde2005logarithmic}.} 
\label{fig:2}
\end{figure}

Starting with a small group of randomly linked nodes, we can simulate the tinkered evolution of a software network via the following copying rules. We grow the network one node at a time. A new node is linked to $m$ randomly chosen target nodes with probability $p$, as well as all ancestor nodes of each target with probability $q$ (see fig.~\ref{fig:2}c). This process continues until a network of size $N$ is formed.  The associated mathematical model for the total number of connections $L(N)$ is:
\begin{equation*}
{{dL}\over{dN}} = mp + mq {{L}\over{N}}.
\end{equation*}
The solution of this equation defines three possible behaviors for the average degree $\left <K \right> = L/N$ depending on the parameter values. For $mp<1$, the network maintains a constant average degree:
\begin{equation*}
\left <K \right> \sim {{mp}\over{1-mq}}.
\end{equation*}
While $mp>1$ causes faster growth, 
\begin{equation*}
\left <K \right> \sim N^{mq-1}
\end{equation*}
resulting in a fully linked network. Software networks seem to be situated precisely at a critical point $mp=1$, which serves as a boundary between a lack of growth in average degree and a highly interconnected state. At this point, average degree grows logarithmically with the number of nodes (see fig.~\ref{fig:2}d): 
\begin{equation*}
\left <K \right> \sim \log(N).
\end{equation*}

In addition, the in-degree distribution follows the scaling equation $P(k)\sim k^2$ with an exponent that is independent of parameter values, whereas the out-degree distribution decays exponentially~\cite{krapivsky2005gnc}. 

The above exposes how some statistical properties of software networks are independent of particular selection forces while some are influenced by cultural evolutionary mechanisms. Although software structure is created by humans, it cannot be free of the constraints that come with evolutionary processes. For many years, we have argued that engineering differs from biology in terms of our ability to forecast the outcomes of our actions. However, when designing complex software systems, continuous experimentation is needed to determine what is correct, and it is very difficult (if not impossible) to anticipate the necessarty changes ahead of time~\cite{soergel2014rampant}. This scenario resembles organismic evolution, in which random changes (mutations) are filtered according to the selection pressure created by environmental conditions. However, a deeper parallel between organisms and software may extend beyond the driving role of selection (either natural or directed by human goals), as both systems have traits influenced by neutral evolution~\cite{lynch2007frailty}. In other words, both organisms and software exhibit complexity that cannot be solely attributed to adaptation.  

\section*{Eco-evolutionary Feedbacks in Software}

Besides internal constraints and for-free complexity afforded by tinkering, software displays clear signatures of eco-evolutionary feedbacks underlying its evolution. In this section we will explore some key structural and temporal patterns relating to various kinds of ecological interactions among technological agents. This classification is necessarily a simplification of a complex hierarchy of firms, programming languages, individual programmers and code. Quantifying diversity and structure across scales is crucial for understanding this complex web of interactions, reflecting a balance between cultural mechanisms of innovation, reinforcement, and replacement~\cite{richerson2008not}.  

\subsection*{Parasitism}

Programming languages, which are comparable to human languages but designed for algorithm execution, are a main driver of software evolution. Programming languages are a set of rules that govern grammar and semantics used to write software, as well as a ``runtime environment," which includes a compiler and libraries for translating the ``high-level" code into actual hardware instructions. Because of their evolving traits, transmission via social learning, and selective spread of certain dialects or variants over others, natural and artificial languages are both paradigmatic examples of cultural evolution~\cite{mace2005phylogenetic, valverde2015punctuated}. 

Language differs from organism evolution in that it lacks a cultural genome, making it difficult to apply traditional genetic methods to reconstruct its evolution~\cite{mace2005phylogenetic}. However, we can still infer information flows using influence or similarity between languages~\cite{valverde2015punctuated}. Furthermore, the links between ancestor and descendant languages reveal ecological interactions when these coexist in the same technological niche, which may include antagonistic relationships. For example, TypeScript and Clojure are ``parasitic'' programming languages built on top of the popular JavaScript, leveraging its stability, infrastructure, and more importantly, its large user base~\cite{hickey2020history}. Hence, influence links capture both ecological and evolutionary dimensions of cultural evolution.  

The phylogenetic network reconstruction of the evolution of programming languages detects events of rapid diversification~\cite{valverde2015punctuated, valverde2016major} (fig.~\ref{fig:3}a). After 1972, there is a larger rate of linguistic trait recombination (horizontal transmission links), especially between the two major programming language families: imperative (descending from Fortran) and functional (descending from Lisp). In line with the pattern of punctuated equilibria found in the fossil record~\cite{eldredge2015eternal, obrien2024punctuated}, external factors could have accelerated abrupt changes in language evolution--for example, the wide adoption of microprocessor technologies as a democratizing factor in software technologies, as well as opening up new niches. 

Conversely, key features developed in the most prolific languages find their way into unrelated ones, which could be interpreted as purely internally driven. For instance, language diversification may be facilitated via ecological interactions. Parasitic languages, as they evolve to better exploit hosts, have the potential to cause evolutionary changes in their host languages~\cite{hickey2020history} (i.e., prompting an evolutionary arms race between hosts and parasites~\cite{smith2007coevolutionary}). This phenomenon may result in the divergence of languages, with significant differences between those that experience high levels of parasitism and those that do not.

In recent years, large language models (LLMs) have introduced a new layer of parasitism within the software ecosystem. Unlike traditional programming languages, which evolve through direct human intervention, LLMs do not generate new syntax or execution environments. Instead, they "feed" on existing human-generated code, learning from repositories, forums, and software documentation to generate new outputs. This process mirrors the parasitic relationship seen in programming language evolution, where certain languages rely on host environments without contributing novel infrastructure.

A key example of this dynamic is how LLMs interact with knowledge-sharing platforms such as Stack Overflow. Empirical studies show that developers increasingly rely on AI-generated responses, leading to a measurable decline in human contributions~\cite{delRioChanona2024}. In contrast to human-mediated knowledge exchange, which refines and improves best practices, LLMs merely extract patterns from existing datasets. This asymmetric relationship risks reducing the diversity of programming techniques by reinforcing established conventions over novel experimentation.

From an evolutionary perspective, LLMs may be seen as a new form of horizontal knowledge transfer, akin to how parasitic languages introduce foreign syntactic elements into host languages.  LLMs have quickly grown since the dominance of decoder-only architectures like GPT-3 and GPT-4~\cite{yang2024harnessing}. Unlike traditional programming languages, which undergo gradual divergence and recombination, LLMs evolve in a highly centralized manner, controlled by a few major research labs. This shift raises fundamental questions about whether AI-driven code generation will foster continued software innovation or lead to increasing uniformity and stagnation.

\subsection*{Competition}

The adoption and use of programming languages frequently results in the formation of communities~\cite{valverde2007self, schueller2022evolving}, shared practices, and even cultural norms centered on their application. These interactions and collaborations contribute to the social dimension of software, often neglected but crucial in understanding its evolution. Interoperability and ease of dissemination are major driving forces in the development of new programming language features, and allow for large-scale standardization across the software landscape~\cite{valverde2021long}. Moreover, programmers tend to learn those languages that are more useful to them, that is, those that are more widely adopted by their peers and contain larger user bases. 

At a higher level, the necessity for individual programmers to collaborate translates into a competition by the programming languages, which vie for a limited amount of agents. Cultural transmission models, integrating an ecological perspective on the underlying social fabric and cultural norms can offer some insights on the pattern of diversification and extinction of programming languages~\cite{valverde2015cultural}. The empirical popularity of coexisting programming languages follows a discrete generalized beta distribution (DGBD)~\cite{martinez2009universality} (inset, fig.~\ref{fig:3}b), linking a language frequency ($f$) to its rank ($r$) :

\begin{equation*}
f(r) = {{A}\over{r^a}} (R+1-r)^b,
\end{equation*}
where $R$ is the maximum rank value, $A$ is a normalization constant and $a, b$ the two fitting exponents (here $a \sim 1.44$, $b \sim 0.46$). Mart\'inez-Mekler {\em et al}. proposed a general model for the DGBD in various cultural systems with an stochastic growth model incorporating two opposed forces: reinforcement and removal of cultural variants~\cite{martinez2009universality}. These two processes map to the fitting parameters of DGBD ($a,b$ respectively), creating steep, long-tailed popularity distributions when reinforcement dominates ($a>b$), and flatter more ``disordered" distributions in the opposite case. 

An alternative explanation for the DGBD relies on competition between cultural variants~\cite{valverde2015cultural}. This view considers programmers not as agents but as a finite resource, that programming languages ``infect" at the expense of other languages. The growth rate of any given language ($d\rho_i / dt$) depends on its current popularity share ($\rho_i$), following the general set ODEs:
\begin{equation*}
{{d\rho_i}\over{dt}} = \mu_i \rho_i \left (\rho_i-\sum_{j\neq i}\rho_j \right ) -\mu_i \rho_i \phi(\overrightarrow{\rho}),
\end{equation*}
where $\mu_i$ is the intrinsic growth rate of the $i$th language and $\phi(\overrightarrow{\rho})$ stands for the competition function that considers the frequency of the focal language $i$ as well as all putative competitors. This system can be made spatially explicit (inset, fig.~\ref{fig:3}b), adding new terms for language transmission according to local density and the discovery of new languages. These rules represent a simple cultural diffusion process, in which variant fitness is defined by its popularity share. In ecological terms, the system dynamics are driven by competitive exclusion, converging towards $n$ languages shared by all programmers (figure~\ref{fig:3}b). The number of final languages $n$ is the precisely the size of the niche, that is, how many different languages are programmers typically proficient in.

\subsection*{Cooperation}

Antagonism is not the only mode of ecological interaction among technological agents that can generate diversity. Cooperation—through coordination strategies and the maintenance of public goods—can increase diversity and complexity, while failure to cooperate can lead to a collapse in both. Understanding the interplay between reward structures, population dynamics, and sociality requires evolutionary models that integrate multiple-choice selection with social learning.

Let's define the probability $\Pi(z)$ of choosing a variant $z \in \{ z_1, z_2, z_3, ... \}$ of a trait as follows: 
\begin{equation*}
\Pi(z) \sim \Phi(z) \times s (p_z),
\end{equation*}
where $\Phi(z)$ is the fitness of the variant (a phenotype-fitness map), $s(p_z)$ is the frequency-dependent selection coefficient,  $p_z = n_z/N$ is the frequency of the trait in the population, $n_z$ is the density of the trait, and $N$ is the number of individuals in the population. This model describes the interaction of different selection pressures in cultural evolution. Directional selection implies that fitness $\Phi(z)= e^{\beta z}$ rises exponentially with trait value $z$, mediated by pay-off transparency ($\beta$)~\cite{lande1983response}.  Conversely, frequency-dependent selection (FDS) emphasizes the social dimension of cultural evolution by increasing the fitness of a trait as it becomes more common~\cite{ayala1974frequency}. FDS follows the scaling relationship  $s(p_z)\sim p_z^J$, capturing negative $(J>0)$, positive $(J>0)$, and frequency-independent ($J=0$) selection, where the exponent $J$ quantifies social imitation~\cite{vidiella2022cultural}. 

This model, which includes directional selection and FDS, may help explain the diversity of behaviors seen in software development even in the absence of external factors. For example, the model predicts an endogenous origin of punctuated evolution linked to positive FDS~\cite{obrien2024punctuated}. In this case, the transition from stasis to punctuation is not due to external shocks; rather, it's the result of accumulating many gradual advances that, over time, overcome adoption barriers associated with conformity bias and imitation~\cite{vidiella2022cultural}. The height of these barriers is defined by the minimal size $\Delta z= z_1- z_0$ of an adaptive step: 
\begin{equation*}
\Delta z \ge {{J-1}\over{\beta}} \log(N),
\end{equation*}
where $z_1$ is the mean trait after the step, and $z_0$ is the main trait before the step.  Although imitation can be beneficial to speed up cultural evolution~\cite{richerson2008not}, no new traits can be easily discovered with exceedingly large $J / \beta$ ratios; instead, individuals only reinforce the popularity of current information, regardless of how obsolete it may be.

The balance between imitation and innovation also shapes software resilience and adaptability. While code reuse reduces development costs, excessive ``copy-paste" programming can lead to bug propagation~\cite{li2006cp}, higher maintenance costs~\cite{eick2001does}, unnecessary complexity~\cite{valverde2017breakdown}, and code ``bloating"~\cite{langdon1998fitness}, particularly in novelty-driven markets~\cite{dewett2007innovators}. 
However, history shows that over-exploitation of successful but repetitive strategies can contribute to cultural stagnation and collapse. Lehman and Stanley argue that abandoning fixed design objectives in favor of novelty prevents stagnation~\cite{lehman2011abandoning}, yet an analysis of Atari 2600 video game development (1977-1982) suggests the opposite pattern~\cite{duran2022dilution}. Before the video game industry crash of 1983, rampant code reuse led to declining lexical diversity and complexity (Figure~\ref{fig:3}c). 

This trend, quantified through the Kolmogorov--Chaitin complexity~\cite{chaitin1966length} of code structures (see fig.~\ref{fig:1}f), is typically uncomputable; however, it can be approximated via the Coding Theorem~\cite{zenil2018decomposition}: 
\begin{equation*}
K(x)=-\log m(x)+O(1),
\end{equation*}
where $m(x)$ is the probability that string $x$ is produced by a random program, and $O(1)$ is a constant independent of $x$~\cite{zenil2018decomposition}. Frequent strings have lower complexity, meaning widespread imitation can systematically reduce software diversity. In anticipation of the impending collapse, an increasing number of Atari 2600 games were produced as clones, detectable by the complexity analysis of their code. The Commodore VIC-20 followed a similar rise-and-fall trajectory in game production, but its software maintained complexity (see fig.~\ref{fig:3}d). Unlike Atari's imitation-driven collapse, the VIC-20 became obsolete through technological succession~\cite{strotz2023end}, replaced by the more advanced Commodore 64. 

A major challenge to cooperative knowledge-sharing in software development is the risk of self-referential learning loops introduced by AI-generated content. Recent studies suggest that when LLMs are trained on their own generated outputs, they begin to lose diversity and accuracy, a phenomenon known as model collapse~\cite{shumailov2024ai}. Just as cultural evolution depends on a balance between imitation and genuine innovation, software ecosystems require a continuous influx of diverse human contributions to remain adaptive. However, as AI-generated code and documentation proliferate, newer models may increasingly be trained on synthetic data, reinforcing established patterns rather than fostering novel solutions. Over time, this feedback loop may erode the adaptive capacity of software development, making it more fragile to external shocks and less capable of generating truly innovative solutions.

This mirrors past cultural collapses where over-reliance on imitation led to stagnation, such as the Atari 2600 software ecosystem, where excessive code reuse ultimately contributed to the industry’s decline~\cite{duran2022dilution}. If AI-generated content dominates software development practices, a similar fate may await software engineering, where innovation is replaced by the blind reinforcement of existing conventions. Unlike human-driven cooperation, which refines and expands knowledge, AI-mediated learning risks amplifying its own biases, accelerating the homogenization of programming practices rather than promoting diversity.

\begin{figure}[htbp]
    \centering
\includegraphics[height=.9\textwidth, angle=90]{Figures/FIG3_Completa_v3.pdf}
    \caption{{\bf Long-term evolutionary trends in software systems}. {\bf (a)} The phylogenetic tree of programming languages exhibits a punctuated pattern~\cite{valverde2015punctuated} with substantial variations in the number of descendants on distinct branches (link thickness). Key innovations in programming languages such as C and Java led to evolutionary radiations. More recently, languages coming from the two primary language families, ...}  
\label{fig:3}
\end{figure}

\begin{figure}[ht!]
    \ContinuedFloat
    \caption[]{
    ...  namely functional (blue) and imperative (red), are converging due to increasing trait recombination rates (yellow links). {\bf (b)} An ecological perspective on technological competition forecasts the collapse of programming language diversity (red squares), using simple rules of competitive exclusion and an upper bound to the number of known languages per user~\cite{valverde2015cultural}. A spatially explicit population dynamics model (top spheres, darker colors represent higher language diversity) reproduces the log-linear rank-popularity relationship (inset), with good agreement between the model (blue circles) and empirical (yellow circles) popularity. {\bf (c)} Average yearly Kolmogorov--Chaitin complexity of Atari 2600 video games (blue) and the number of games released each year (red). {\bf (d)} Equivalent plot for VIC-20 code complexity (blue), as well as productivity (red). The significant decrease in code complexity precedes the economic collapse of 1983, suggesting a causal link between innovation and product value. Similar trends in lexical diversity and code compressibility indicate significant increases in code redundancy due to imitation~\cite{duran2022dilution}.}
\end{figure}

\subsection*{Symbiosis}

Symbiotic interactions in software emerge when two systems merge, typically when a larger project absorbs a smaller or discontinued one, or when a project integrates external codebases. These interactions resemble biological symbiosis, where species co-evolve in mutually dependent relationships. In software, symbiosis can take different forms. In some cases, a dominant system extracts functionality from another without major modifications, resembling parasitism. In others, a project may benefit from another without significantly altering it, a dynamic closer to commensalism. However, the most productive forms of symbiosis are mutualistic, where two systems evolve together, improving performance, stability, or adaptability.

A common example of software symbiosis occurs when an open-source project incorporates a third-party library, introducing complex co-evolutionary dynamics. Initially, the host system exhibits a transition from endogenous (internally driven) to exogenous (externally driven) scaling dynamics, while the imported library remains relatively inactive~\cite{valverde2005logarithmic, valverde2007crossover}. As the library becomes fully integrated, it accumulates modifications and begins to co-evolve with the host system, resulting in coupled development cycles. These interactions shape the long-term trajectory of software evolution by determining how different components interact and adapt to changes in their environment.

Software activity fluctuations reveal a central tension between internal maintenance tasks and external requirements. At the individual developer level, non-homogeneous activity patterns suggest self-organization, indicating that the social structure of development teams may influence long-term software evolution. Specifically, the distribution of time delays between consecutive modifications in a software project follows a Weibull distribution:
\begin{equation*}
P_>(T) \sim e^{-(T / \left < T \right >)^\alpha},
\end{equation*}
where $\alpha \sim 0.6$~\cite{challet2008fat}.
This distribution differs from the pure power-law behavior observed in financial markets and other technical systems~\cite{eisler2006size}. The presence of non-Poissonian statistics suggests that software and financial systems share a key feature: strong agent interactions that induce long-term memory effects. In software, this implies that past development decisions influence future modifications, leading to path-dependent evolution~\cite{sole2013evolutionary}. This persistence of historical structures is a crucial factor in understanding how software ecosystems remain stable or adapt over time. Just as symbiotic relationships in biological systems can shape evolutionary outcomes, software symbiosis influences the long-term sustainability and resilience of technological artifacts.

\section*{Discussion}
Understanding the evolution of software requires a holistic approach that incorporates both neutral processes and an ecological perspective.  Common network properties like scale-free behavior and the heterogeneous distribution of motifs evidence the widespread reuse of selected components. Contrasting these empirical observations with theoretical predictions can reveal underlying feedback loops between software adaptation and environmental constraints. When software components interact with human agents, they create an artificial ecosystem characterized by elements such as parasitism, competition, collaboration, and symbiosis. Similar to natural ecosystems, cooperation and mutualism among software components can enhance the tempo and mode of software innovation. For example, collaborative development processes often lead to faster and more creative solutions~\cite{gloor2006swarm}.

Large language models (LLMs) introduce a new dynamic to this artificial ecosystem. LLMs offer developers quick access to technical information while also introducing new selective pressures that shape software evolution. A key concern is that their widespread use may disrupt the traditional balance between cooperation, imitation, and innovation. By providing immediate AI-generated solutions, LLMs reduce the need for developers to engage in collaborative problem-solving and knowledge refinement~\cite{delRioChanona2024}. If this trend continues, it could alter the cultural transmission mechanisms that have historically governed software development, fundamentally reshaping how knowledge and innovation propagate within the industry. Instead of an ecosystem driven by the exchange of novel solutions, software innovation may increasingly rely on imitation and AI-reinforced conventions, accelerating a shift toward homogenization. Its effects can already be seen in the prevalence of repetitive coding patterns and the widespread adoption of standardized AI-generated solutions~\cite{wang2023survey}.

This trend is consistent with the dilution of expertise hypothesis, which suggests that when a domain like software development expands and becomes more accessible to a broad, non-expert audience, there is a danger of over-reliance on imitation, which is harmful to innovation~\cite{duran2022dilution}. In the past, software diversity has been sustained by a dynamic equilibrium between exploration and exploitation, where developers refine solutions through iterative feedback. LLMs, by making common solutions immediately accessible, may weaken the incentives for deep technical exploration. Over time, this could lead to a narrowing of software diversity, as developers converge on AI-suggested implementations rather than independently experimenting with alternatives.

The above suggests that software complexity and innovation cannot be reduced to our ability to act purposefully on the environment. Instead, software is an emergent property shaped by both directed changes and eco-evolutionary feedbacks. Software, while a machine-based representation of a mathematical algorithm, also serves as a non-material support for social interactions that foster innovation and diversity. The dual nature of software, serving as both a technical product and a social phenomenon, significantly influences cultural evolution, despite being understudied. With the rising prominence of AI-mediated knowledge production, a crucial question arises: will LLMs accelerate software evolution, or will they precipitate a cultural collapse where imitation surpasses innovation?

\begin{acknowledgement}
The authors thank François Lafond and Doyne Farmer for their kind invitation to participate in this book. S.V. acknowledges his colleagues in the video game industry, in particular Jos\'e Paredes, Alex Rodriguez, Francisco Jos\'e Garc\'ia-Ojalvo, Jos\'e Andreu, and Eduardo Arancibia, for all the shared lessons and conversations over the years. S.V. is supported by the Spanish Ministry of Science and Innovation through the State Research Agency~(AEI), grant PID2020-117822GB-I00 /AEI /10.13039/501100011033. 
S.D-N. is supported by the Beatriu de Pinós postdoctoral programme, from the Office of the General Secretary of Research and Universities and the Ministry of Research and Univertisites (2019 BP 00206) and the support of the Marie Sklodowska-Curie COFUND (BP3 contract no. 801370) of the H2020 programme. 
\end{acknowledgement}
\bibliographystyle{unsrtnat}
\bibliography{bibliography}

\end{document}